\documentclass[
amsmath, %
floatfix, %
twocolumn, %
reprint, %
prx, %
aps, %
citeautoscript, %
longbibliography, %
final, %
]{revtex4-2}
\renewcommand{\thispagestyle}[1]{}
\usepackage[]{newtxtext} 
\usepackage[subscriptcorrection,nosymbolsc,smallerops,bigdelims]{newtxmath}
\DeclareMathAlphabet{\mathcal}{OMS}{cmsy}{m}{n}
\DeclareMathAlphabet{\mathbcal}{OMS}{cmsy}{b}{n}
\usepackage{bm}

\usepackage[low-sup]{subdepth}

\usepackage[utf8]{inputenc}
\usepackage[T1]{fontenc}

\usepackage[]{graphicx}
\usepackage{latexsym}
\usepackage{color}
\usepackage{mathtools}
\usepackage[section]{placeins}
\usepackage[]{xcolor}
\usepackage{bm}
\usepackage{bbold}
\usepackage{multirow}
\usepackage{siunitx}
\usepackage{hyperref}
\hypersetup{
	colorlinks,
	linkcolor={blue!90!black},
	citecolor={blue!90!black},
	urlcolor=	{blue!90!black}
}
\let\oldeqref\eqref
\renewcommand*{\eqref}[1]{%
	\hyperref[#1]{\oldeqref{#1}}%
}

\usepackage{floatrow}
\usepackage{placeins}

\newcommand{\mr}[1]{\mathrm{#1}}

\mathchardef\mhyphen="2D

\DeclarePairedDelimiter\aver{\langle}{\rangle}
\DeclarePairedDelimiterX{\comm}[2]{\lbrack}{\rbrack}{#1, #2}
\DeclarePairedDelimiter\ket{\lvert}{\rangle}

\DeclarePairedDelimiterX{\braket}[2]{\langle}{\rangle}{#1\delimsize\vert #2}
\DeclarePairedDelimiterX{\ketbra}[2]{\rvert}{\lvert}{#1 \delimsize\rangle\!\delimsize\langle #2}
\DeclarePairedDelimiterX{\matrixel}[3]{\langle}{\rangle}{#1 \delimsize\vert #2 \delimsize\vert #3}

\newcommand{\hc}{\mr{H.c.}}

\newcommand{\eqnref}[1]{Eq.~\eqref{#1}}

\newcommand{\figref}[1]{Fig.~\ref{#1}}
\newcommand{\subfigref}[2]{Fig.~\hyperref[#1]{\ref*{#1}(#2)}}
\newcommand{\subfigsref}[3]{Figs.~\hyperref[#1]{\ref*{#1}(#2)}-\hyperref[#1]{\ref*{#1}(#3)}}

\newcommand{\asubfigref}[2]{Figure~\hyperref[#1]{\ref*{#1}(#2)}}
\newcommand{\asubfigsref}[3]{Figures~\hyperref[#1]{\ref*{#1}(#2)}-\hyperref[#1]{\ref*{#1}(#3)}}

\usepackage[low-sup]{subdepth}
\usepackage{ragged2e}

\definecolor{cbred}{HTML}{e31a1c}
\definecolor{cbgreen}{HTML}{33a02c}
\definecolor{cbblue}{HTML}{176aa7}
\definecolor{cborange}{HTML}{ff7f00}
\definecolor{cbviolet}{HTML}{6a3d9a}
\definecolor{cbbrown}{HTML}{b15928}

\definecolor{cblred}{HTML}{fb9a99}
\definecolor{cblgreen}{HTML}{b2df8a}
\definecolor{cblblue}{HTML}{a6cee3}
\definecolor{cblorange}{HTML}{fdbf6f}
\definecolor{cblviolet}{HTML}{cab2d6}
\definecolor{cblbrown}{HTML}{ffff99}

\makeatletter
\newcommand{\doublewidetilde}[1]{%
  \mathpalette\doublewidetilde@{#1}%
}
\newcommand{\doublewidetilde@}[2]{%
  \ooalign{%
    $\m@th#1\widetilde{#2}$\cr
    \raise.35ex\hbox{$\m@th#1\kern1pt\widetilde{\phantom{#2}}$}\cr
  }%
}
\makeatother

\usepackage[activate={true,nocompatibility},final,tracking=alltext,kerning=true,spacing=true,protrusion=true,factor=1080,stretch=5,shrink=9,selected=true ,letterspace=-0]{microtype}
\usepackage{orcidlink}

\begin{document}

\title{Higher-harmonic acoustic control of laser-dressed quantum-dot states at sub-Rabi frequencies}

\author{Mateusz Kuniej\,\orcidlink{0000-0001-5476-4856}}
\affiliation{Institute of Theoretical Physics, Wroc\l{}aw University of Science and Technology, 50-370 Wroc\l{}aw, Poland}

\author{Pawe{\l} Machnikowski\,\orcidlink{0000-0003-0349-1725}}
\affiliation{Institute of Theoretical Physics, Wroc\l{}aw University of Science and Technology, 50-370 Wroc\l{}aw, Poland}

\author{Micha{\l} Gawe{\l}czyk\,\orcidlink{0000-0003-2299-140X}}
\affiliation{Institute of Theoretical Physics, Wroc\l{}aw University of Science and Technology, 50-370 Wroc\l{}aw, Poland}

\begin{abstract}
    Acoustic control and coupling of quantum systems via phonons can enable miniaturized on-chip quantum devices. Optically active quantum dots (QDs) are essential for such platforms, yet they have long lacked direct acoustic transitions between charge states. The recently proposed hybrid acousto-optical swing-up scheme introduces such high-fidelity transitions but, as typical for coherent control, requires driving fields matching the transition energy and has been proposed for sub-THz phonon frequencies, limiting practical implementations. Here, we overcome this limitation by exploiting higher-harmonic-assisted processes arising from strain-induced modulation of the optical transition energy. This parametric modulation of the optically dressed splitting produces multi-phonon-like resonances when a harmonic of the mechanical modulation matches the generalized Rabi frequency. An effective model reveals a hierarchy of harmonic couplings, while our geometric interpretation shows how modulated dressed splitting enables even harmonics that are otherwise symmetry-forbidden.
    We predict faithful state preparation with an acoustic frequency being only a fraction of the splitting, specifically 42~GHz for a 0.341~THz splitting, bridging accessible acoustic frequencies and THz energy scales. This establishes control principles that separate optical energy delivery from coherent acoustic control. A non-Markovian calculation of phonon-induced decoherence indicates high state-preparation fidelities comparable to one-phonon and all-optical schemes. Since the same interaction structure arises for a quantized acoustic field, our results provide a foundation for multi-phonon processes in QD-phononic-resonator systems, including entanglement, state transfer, and optical preparation of nonclassical mechanical states.
\end{abstract}

\maketitle

\section{Introduction}
Acoustic waves \cite{Wang2018, Choquer2022} couple to all solid-state systems, classical or quantum. This property enables various applications, including quantum technologies \cite{Delsing2019, Roy2011, Fu2017}. Recent advances in generating coherent acoustic waves \cite{Nysten2020, Choquer2022} combined with the development of phononic structures \cite{Preuss2022, Roth2023} and the fabrication of surface acoustic wave cavities \cite{DeCrescent2022}, may enable connecting diverse quantum systems into multi-component hybrid architectures \cite{Kurizki2015} through a universal acoustic quantum bus \cite{Schuetz2015,Dumur2021} and controlling various systems acoustically \cite{Wang2024, Dietz2023, DeCrescent2024, Kuniej2025,Kuniej2026}, thus igniting the field of quantum acoustodynamics~\cite{Chu2018, Satzinger2018}.

Optically active quantum dots (QDs) may serve as charge and spin qubits \cite{Loss1998} and high-quality quantum emitters \cite{Michler2000, Dusanowski2023, Tomm2021}, providing single \cite{Senellart2017, Musial2020} entangled \cite{Liu2019, Hafenbrak2007}, and cluster photon states \cite{Cogan2023, Economou2010, Lindner2009, Michaels2021}. Several QD state preparation methods, including resonant excitation \cite{Stievater2001, Kamada2001}, two-photon excitation \cite{Stufler2006, Machnikowski2008}, and phonon-assisted processes \cite{Weiler2012, Glassl2013}, have been proposed and realized. A recent promising all-optical approach utilizes the beating of two detuned optical pulses. It moves the optical qubit to the opposite pole of the Bloch sphere along a spiral defined by two tilted rotation axes \cite{Bracht2021, Karli2022}, and also applies to other quantum emitters \cite{Torun2026}.

Although carrier-deformation coupling in QDs cannot drive optically allowed transitions, optomechanical wave mixing and modulated resonance-fluorescence spectral dynamics have been explored \cite{Weiss2021, Wigger2021, Groll2025, Zhan2025}, with recent surface-acoustic-wave designs allowing selective quantum-dot modulation \cite{Zhu2025}. To excite QD charge states using acoustic waves, one can exploit mechanically assisted photon scattering \cite{DeCrescent2024} or the recently proposed acousto-optical swing-up scheme, involving direct acoustic modulation of the detuning \cite{Kuniej2025}. We focus on the latter scheme, which can be nearly phonon-decoherence-free and enables acoustic or acoustically gated optical control of quantum emitters, with potential to generate QD-phonon entanglement and achieve acoustic state transfer. However, this method encounters a bottleneck due to the challenging necessary phonon energy in the sub-THz range.

Here, we propose a way to overcome the limited availability of acoustic drive frequencies, thereby significantly expanding the method's application potential. We show that acousto-optical swing-up extends naturally to a fractional-frequency regime enabled by multi-phonon-like processes involving higher acoustic harmonics. These harmonics arise effectively from periodic phase modulation \cite{Steane1995} of excitonic optical transitions \cite{Imany2022, Villa2017}, enabling parametric multi-phonon-like resonances with the laser-dressed Rabi splitting even when the modulation is provided by a classical acoustic wave. Our study identifies the resonance conditions, optimal modulation strengths, and decoherence constraints for such higher-harmonic parametric control.

By calculating the driven evolution and decoherence within a non-Markovian framework, we show that the scheme enables optically gated acoustic control of the QD charge state, including high-fidelity preparation of exciton and biexciton states in application-relevant local-droplet-etched GaAs/AlGaAs QDs \cite{daSilva2021}. Moreover, the fidelity is almost independent of the harmonic order, and we show that at least an order of magnitude in frequency can be bridged using a ten-phonon-like process to drive sub-THz dynamics. While we focus on the regime where the acoustic field can be treated as a classical drive, the theory is structurally identical to that for a quantized acoustic mode in the large-coherent-amplitude limit (we thus further use the term `multi-phonon' where higher acoustic harmonics are involved). This establishes mechanically programmable coherent control of QD states and constitutes a necessary first step towards fully quantum multi-phonon dynamics, where an engineered mechanical resonator could support nonclassical phonon states generated and manipulated via the same multi-phonon-type couplings in a QD.

\begin{figure}[tb!]
    \centering
    \includegraphics[width=0.9\linewidth]{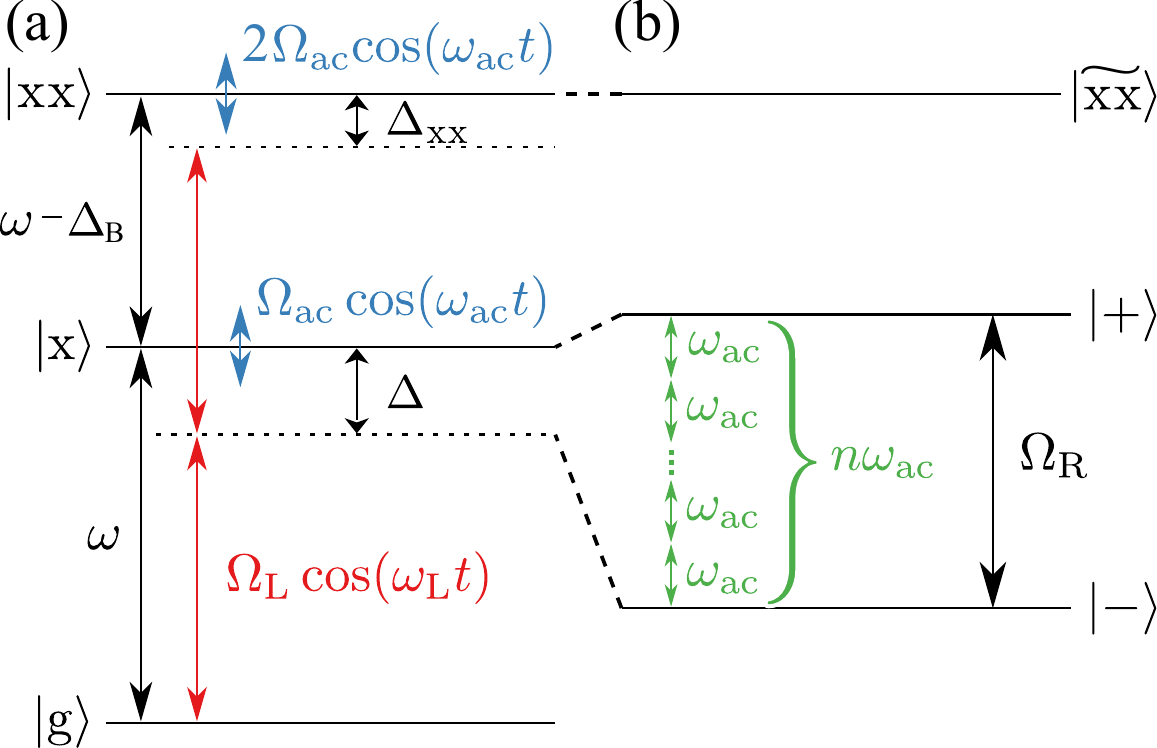}
    \caption{(a) Schematic energy structure of the three-level system with external fields. Laser (red arrows) couples charge states $\ket{\mathrm{g}}$ and $\ket{\mathrm{x}}$ corresponding to empty QD and exciton state with detuning $\Delta$, while it is detuned from the two-photon transition to the biexciton state $\ket{\mathrm{xx}}$ by $\Delta_{\mathrm{xx}}$. Acoustic modulation of states $\ket{\mathrm{x}}$ and $\ket{\mathrm{xx}}$ is marked with the blue arrows. (b) States of the system dressed by light; exemplary case of strong $\ket{\mathrm{g}}\leftrightarrow\ket{\mathrm{x}}$ mixing into $\ket{+}$ and $\ket{-}$, while $\ket{\widetilde{\mathrm{xx}}}$ is a weakly perturbed $\ket{\mathrm{xx}}$ state. Acoustic field energy is marked with green arrows, and its multiple $n\omega_{\mathrm{ac}}$ corresponds to the $\ket{+}$--$\ket{-}$ splitting.}
    \label{fig:system}
\end{figure}

\section{Theory}
\subsection{System and model}
\asubfigref{fig:system}{a} presents the schematic level diagram of the system comprising a semiconductor QD coupled to an off-resonant linearly polarized laser, with detuning $\Delta$ from the fundamental transition. Appropriate linear polarization selects one optical path for transitions between the ground state $\ket{\mathrm{g}}$ and one of the bright exciton states $\ket{\mathrm{x}}$ (we neglect the other), and between $\ket{\mathrm{x}}$ and the biexciton $\ket{\mathrm{xx}}$, with splittings $\hbar\omega$ and $\hbar(\omega - \Delta_{\mathrm{B}})$, respectively, where $\hbar\Delta_{\mathrm{B}}$ is the biexciton binding energy. The system is additionally subject to an acoustic wave with frequency $\omega_{\mathrm{ac}}$, which modulates the exciton and, approximately twice as strongly, the biexciton energy. Although the carrier-deformation coupling in QDs is diagonal in charge states, acousto-optical control of charge states was predicted and demonstrated \cite{Kuniej2025, DeCrescent2024}. Here, we extend the acousto-optical swing-up scheme to enable multi-phonon transitions between QD states. This approach fundamentally differs from the all-optical swing-up scheme, in which QD excitation involves photon scattering between different laser modes \cite{Vannucci2024, Richter2025}, whereas here population transfer is mediated by multi-phonon resonances with integer multiples of the acoustic frequency.

The Hamiltonian of the system with optical coupling and acoustic modulation, in the rotating wave approximation and frame rotating at the laser frequency $\omega_{\mathrm{L}}$, is~\cite{Kuniej2025}
\begin{equation}
    \begin{split}
        H(t) = &-\hbar\Delta\ketbra{\mathrm{x}}{\mathrm{x}} - \hbar\Delta_{\mathrm{xx}}\ketbra{\mathrm{xx}}{\mathrm{xx}} \\
               & + \frac{1}{2}\hbar\Omega_{\mathrm{L}}(t)(\ketbra{\mathrm{x}}{\mathrm{g}} + \ketbra{\mathrm{xx}}{\mathrm{x}} + \hc) \\
               & + \hbar\Omega_{\mathrm{ac}}(t)\cos(\omega_{\mathrm{ac}}t)\left(\ketbra{\mathrm{x}}{\mathrm{x}} + 2\ketbra{\mathrm{xx}}{\mathrm{xx}}\right) \\
             \equiv & \phantom{+}H_0 + H_{\mathrm{L}} + H_{\mathrm{ac}},
    \end{split}
    \label{eq:threeLevelSystemHamiltonianRWA}
\end{equation}
where $\Delta = \omega_{\mathrm{L}} - \omega$ and $\Delta_{\mathrm{xx}} = 2\Delta + \Delta_{\mathrm{B}}$ are the detunings from the ground state-exciton and two-photon ground state-biexciton transitions, respectively. $\Omega_{\mathrm{L}}(t)$ and $\Omega_{\mathrm{ac}}(t)$ are the field envelopes with maximal values (plateau) $A_{\mathrm{L}}$ and $A_{\mathrm{ac}}$ with subscripts indicating the optical (L) and acoustic (ac) fields.
To obtain an exact evolution, we solve the Liouville-von Neumann equation $\dot{\rho}(t)=(-i/\hbar)[H(t),\rho(t)]$ for the density matrix $\rho$ with the initial state $\rho(0)=\ketbra{\mathrm{g}}{\mathrm{g}}$.
When studying decoherence, we treat the QD as an open system in contact with a phonon bath, with the total Hamiltonian
\begin{equation}
    H_{\mathrm{full}} = H + H_{\mathrm{ph}}+V,
\end{equation}
where $H_{\mathrm{ph}}=\hbar\sum_{\bm{q}\lambda}\omega_{q\lambda}b_{\bm{q}\lambda}^\dagger b_{\bm{q}\lambda}$ is the free-phonon Hamiltonian with $b_{\bm{q}\lambda}^\dagger$ creating a phonon with wavevector $\bm{q}$ from the $\lambda$ branch (acoustic) with frequency $\omega_{q\lambda}$, and
\begin{equation}
    V = \left( \ketbra{\mathrm{x}}{\mathrm{x}}+2\ketbra{\mathrm{xx}}{\mathrm{xx}}\right)\otimes\sum_{\bm{q}\lambda}g_{\mathrm{x}}^{(\bm{q}\lambda)}\left(b_{\bm{q}\lambda}+b_{\bm{-q}\lambda}^\dagger\right)
\end{equation}
is the carrier-phonon coupling with strength $g_{i}^{(\bm{q}\lambda)}$ for a charge state $i$, and we have used $g_{\mathrm{g}}^{(\bm{q}\lambda)}\!=0$, $g_{\mathrm{xx}}^{(\bm{q}\lambda)}\!\approx2g_{\mathrm{x}}^{(\bm{q}\lambda)}$.
We evaluate $g_{\mathrm{x}}^{(\bm{q}\lambda)}$ and the phonon-induced coherence loss for a typical GaAs/AlGaAs QD in a non-Markovian approach \cite{Alicki2002, Alicki2004} adopted previously \cite{Kuniej2025} and described in the Supplementary Information~\cite{supplement}, where the material parameters and details of the QD-phonon coupling are also given.

To analyze the evolution and derive an analytical formula for multi-phonon processes, we first neglect the biexciton state, simplifying the problem to a two-level system. Next, we diagonalize $H_0 + H_{\mathrm{L}}$. For a slowly varying laser envelope, this is done at each moment, yielding instantaneous dressed eigenstates $\ket{\pm}$ as combinations of $\ket{\mathrm{g}}$ and $\ket{\mathrm{x}}$ with a time-dependent mixing angle $\theta(t) = \tan^{-1}(\Omega_{\mathrm{L}}(t)/\Delta)$ and a splitting corresponding to the optical Rabi frequency, $E_+(t) - E_-(t) \equiv \hbar\Omega_{\mathrm{R}}(t)$. In this basis, the acoustic field coupling is no longer diagonal and takes the form
\begin{align}\label{eq:acousticHamiltonianTwoLevel}
        H_{\mathrm{ac}}(t) = {}& \hbar\Omega_{\mathrm{ac}}(t)\cos(\omega_{\mathrm{ac}}t) \notag \\
        &{}\times\left(\sin^2\!\frac{\theta(t)}{2}\ketbra{+}{+} + \cos^2\!\frac{\theta(t)}{2}\ketbra{-}{-}\right) \\
                             & {}+ \frac{1}{2}\hbar\Omega_{\mathrm{ac}}(t)\cos(\omega_{\mathrm{ac}}t)\sin\theta(t)\left(\ketbra{+}{-} + \ketbra{-}{+}\right). \notag
\end{align}
Alongside the key off-diagonal elements that enable the acoustic driving of transitions when $\omega_{\mathrm{ac}} = \Omega_{\mathrm{R}}$ \cite{Kuniej2025}, the diagonal terms parametrically modulate the energy of dressed states and thereby the Rabi frequency, in addition to its slow variation due to envelopes, 
\begin{equation}
    E_{+}^{(\mathrm{ac})}(t) - E_{-}^{(\mathrm{ac})}(t) = \hbar\Omega_{\mathrm{R}} - \hbar\Omega_{\mathrm{ac}}(t)\cos(\omega_{\mathrm{ac}}t)\cos\theta(t).
\end{equation}

\subsection{Effective Hamiltonian}
Unlike original approaches for all-optical \cite{Bracht2021} and acousto-optical swing-up \cite{Kuniej2025}, we do not neglect the modulation of $\Omega_{\mathrm{R}}$, demonstrating its essential role in parametric acousto-optical control via exchange of integer multiples of the acoustic frequency (multi-phonon processes). Let us assume both envelopes constant for simplicity, $\Omega_{\mathrm{L}}(t), \Omega_{\mathrm{ac}}(t), \theta(t) = \text{const.}$ To highlight multi-phonon dressed-state resonances, we apply an additional unitary transformation, $\widetilde{H}(t) = S(t)H(t)S^{\dagger}(t) + i\hbar\Dot{S}(t)S^{\dagger}(t)$, with respect to the term that drives the $\Omega_{\mathrm{R}}$ modulation, i.e., with
\begin{equation}
    S(t) = e^{i\Omega_{\mathrm{ac}}\!\int_{t_0}^{t}\!\!\mathrm{d}\tau\cos(\omega_{\mathrm{ac}}\tau)\left(\sin^2\frac{\theta}{2}\ketbra{+}{+} + \cos^2\frac{\theta}{2}\ketbra{-}{-}\right)}.
\end{equation}
After integration, we apply the Jacobi-Anger formula, $e^{iA\sin(x)} = \sum_{n}J_n(A)e^{inx}$, to expand $S(t)$ in terms of harmonic functions with fundamental frequency $\omega_{\mathrm{ac}}$. Shifting the energy zero, we obtain the transformed Hamiltonian, 
\begin{equation}
    \begin{split}
        \widetilde{H}(t) = {}& \frac{1}{2}\hbar\Omega_{\mathrm{R}}(\ketbra{+}{+} - \ketbra{-}{-}) + \frac{1}{2}\hbar\Omega_{\mathrm{ac}}\sin\theta \\
                           &\times\sum_{n=-\infty}^{+\infty}\left[J_n(\Omega\cos\theta)e^{in\omega_{\mathrm{ac}}t}\ketbra{+}{-} + \hc\right],
    \end{split}
\end{equation}
where $\Omega = \Omega_{\mathrm{ac}}/\omega_{\mathrm{ac}}$, $J_{n}(x)$ are Bessel functions of the first kind. The evolution is now evident to be driven by the terms $\propto e^{in\omega_{\mathrm{ac}}t}$ with integer $n$, i.e., the harmonics enabled by phase modulation in the system \cite{Steane1995}, corresponding to multi-phonon dressed-state couplings.
Using property $J_{-n}(x)=(-1)^{n}J_{n}(x)$, we express the Hamiltonian as 
\begin{equation}
\widetilde{H}(t) = \hbar\Omega_{\mathrm{R}}\sigma_z + \sum_{n=0}^{\infty} \tilde{J}_n\,
\begin{cases}
\cos(n\omega_{\mathrm{ac}} t)\sigma_x, & n~\text{even},\\
\sin(n\omega_{\mathrm{ac}} t)\sigma_y, & n~\text{odd},
\end{cases}
\end{equation}
where $\tilde{J_n}\equiv \hbar\Omega_{\mathrm{ac}}\sin\theta J_n(\Omega\cos\theta)$, to see that it comprises effective Rabi drives at successive multiples of $\omega_{\mathrm{ac}}$ with different rotation axes for even and odd harmonics. Bringing one of these drives to resonance, $n\omega_{\mathrm{ac}} = \Omega_{\mathrm{R}}$, while keeping others off-resonant, induces multi-phonon transitions between laser-dressed states.
This is apparent after switching to a rotating frame through a unitary transformation with $S(t) = \exp(in\omega_{\text{ac}}t\sigma_z/2)$ and neglecting all nonsecular (oscillating) terms, yielding (for even $n$)
\begin{equation}
    \doublewidetilde{H}(t) \simeq \hbar\left( \Omega_{\mathrm{R}}-n\omega_{\text{ac}} \right) \sigma_z + \frac{1}{2}\tilde{J}_n\sigma_x,
\end{equation}
which is a standard Rabi Hamiltonian (a similar approach applies for odd $n$).
Further numerical calculations demonstrate that this effective model captures the primary processes in the evolution. 

While the illustration uses a two-level model, selective transitions between any pair of states in the full model can be achieved by matching $n\omega_{\mathrm{ac}}$ to the splitting between given dressed states. A similar effect should occur in the all-optical swing-up, where multiples of the beat frequency of the two detuned pulses can be exchanged. However, fundamentally different mechanisms govern the two effects at the quantum level: here, coherent population transfer is mediated by multiple acoustic quanta, whereas the all-optical effect relies on photon scattering between two optical modes \cite{Richter2025, Vannucci2024}.

Intuitively, excitation via this scheme for $n=1$ involves periodic modulation of the inclination of the qubit rotational axis at frequency $\omega_{\mathrm{ac}}=\Omega_{\mathrm{R}}$  \cite{Kuniej2025}. During one optical Rabi period, the state traces semicircles on the Bloch sphere around two axes, ascending more along one semicircle than descending along the other, resembling a spiral path. In the fractional case with $n>1$, the trajectory appears less evident. The explanation seems to fail for even $n$, since a single period should become symmetric: both halves of the period contain identical sets of ascending and descending segments. Consequently, the path should be closed, never reaching the opposite pole. However, the segments are unequal due to the modulation of the Rabi frequency. The rotation about one axis lasts longer than about the other. This asymmetry enables coherent spin rotation despite even $n$. For odd $n$, the evolution over a single Rabi period is intrinsically asymmetric, enabling the scheme to function. In other words, Rabi frequency modulation removes the symmetry constraint that otherwise suppresses even harmonics in the standard Rabi model.

\section{Results}
\subsection{Numerical results}
Having gained intuition, we now study the evolution of the system by numerically solving the Liouville-von~Neumann equation for the full model of \eqnref{eq:threeLevelSystemHamiltonianRWA} and the system initially in the ground state. We take both fields with flat-top envelopes with plateau durations of 0.4~ns (laser) and 0.28~ns (acoustic). Thus, the acoustic pulse is contained within the optical pulse, which corresponds to optically gated acoustic driving. Calculations use feasible envelope switching rates of 0.04~ns (L) and 0.028~ns (ac; single period) and $\hbar\Delta = \hbar\Delta_{\mathrm{xx}} = 0.1$~meV.
We choose $A_{\mathrm{L}} = |\Delta|$ for exciton preparation. For the biexciton, we set $A_{\mathrm{L}} = 0.65$~meV to maintain an optical Rabi frequency similar to that used in exciton preparation.

\begin{figure}[tb!]
    \centering
    \includegraphics[width=1\linewidth]{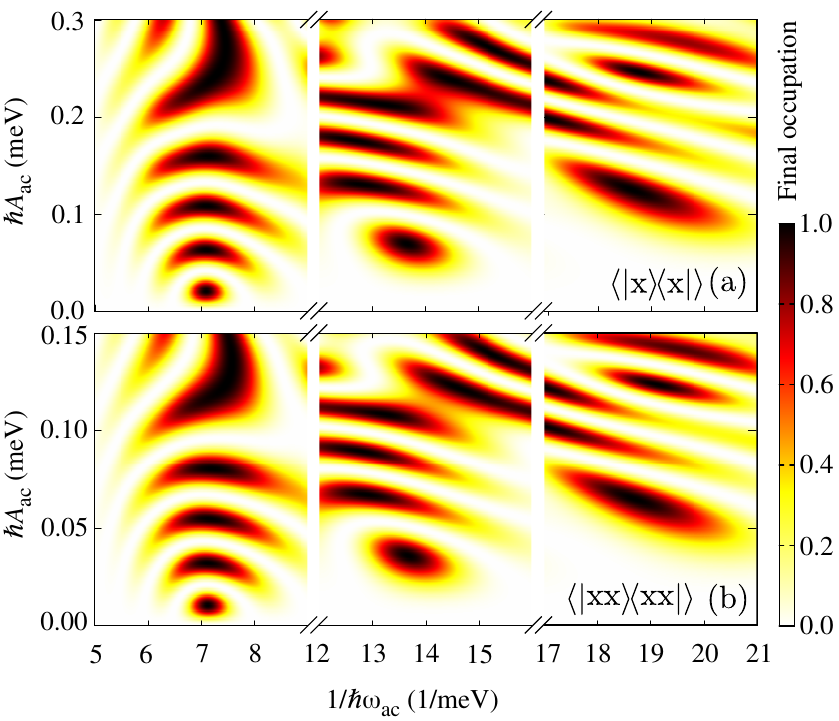}
    \caption{Final occupation of the (a) exciton state, (b) biexciton state as a function of acoustic field amplitude and inverse of its frequency.}
    \label{fig:fractionalMapOccupation}
\end{figure}

\begin{figure*}[tb!]
    \centering
    \includegraphics[width=0.85\linewidth]{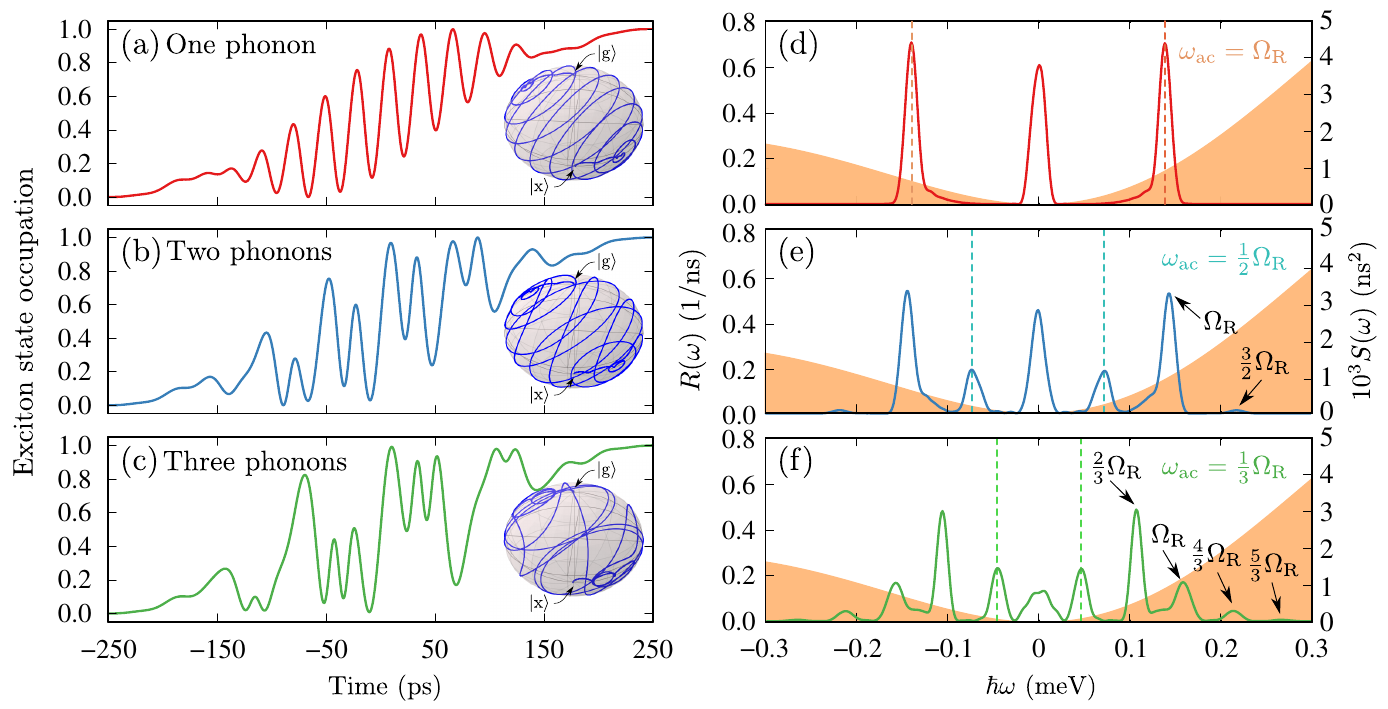}
    \caption{(a)-(c) Evolution of exciton occupation for $\pi$-rotations from \subfigref{fig:fractionalMapOccupation}{a} for (a) one- ($\hbar\omega_{\mathrm{ac}}\approx0.141$~meV,  $\hbar A_{\mathrm{ac}}\approx0.0213$~meV), (b) two- ($\hbar\omega_{\mathrm{ac}}\approx0.0733$~meV,  $\hbar A_{\mathrm{ac}}\approx0.0696$~meV), and (c) three-phonon processes ($\hbar\omega_{\mathrm{ac}}\approx0.0536$~meV,  $\hbar A_{\mathrm{ac}}\approx0.125$~meV). Insets show the state evolution on the $\ket{\text{g}}$-$\ket{\text{x}}$ Bloch sphere. (d)-(f) Nonlinear spectral function $S(\omega)$ for evolution from panels (a)-(c). Filled curves show phonon spectral density $R(\omega)$ calculated for typical GaAs/AlGaAs QDs at $T=4$~K; vertical dashed lines show $\omega_{\mathrm{ac}}$.}
    \label{fig:evolution}
\end{figure*}

\subfigref{fig:fractionalMapOccupation}{a} and \subfigref{fig:fractionalMapOccupation}{b} present the final exciton and biexciton state occupations as functions of acoustic field amplitude and frequency (note the broken axes with irrelevant frequencies omitted). We plot one-, two-, and three-phonon processes (consecutive maxima along the $1/\omega_{\mathrm{ac}}$ axis). The vertical maxima along the $A_{\mathrm{ac}}$ axis correspond to multiple $\pi$-rotations. In both cases, $n=3$ maxima correspond to sufficiently low acoustic frequencies achievable with current interdigital transducers \cite{Delsing2019, Zheng2020, Zhou2023}. \asubfigsref{fig:evolution}{a}{c} show evolutions corresponding to the $\pi$-rotation maxima for one-, two-, and three-phonon processes from \subfigref{fig:fractionalMapOccupation}{a}, respectively. We predict perfect state preparation (neglecting environmental effects) with the oscillation complexity increasing with the acoustic harmonic order. Insets to \subfigsref{fig:evolution}{a}{c} show the qubit evolution on the Bloch sphere.

The evolution is more easily characterized in the spectral domain. For this, we calculate the nonlinear spectral characteristic $S(\omega)=\aver{ Y(\omega) P_{\perp} Y(\omega) }_0$ with $\aver{O}_0 = \text{Tr}O\rho_0$, $Y(\omega)$ the Fourier transform of $\ketbra{\text{x}}{\text{x}}+2\ketbra{\text{xx}}{\text{xx}}$, 
in the interaction picture with respect to $H_0+H_{\mathrm{L}}$,
and $P_{\perp}=\ketbra{\mathrm{x}}{\mathrm{x}}+\ketbra{\mathrm{xx}}{\mathrm{xx}}$ projecting onto the orthogonal complement of the initial carrier state \cite{Roszak2005}. $S(\omega)$ reveals the frequencies present in the evolution of charges, where the factor of 2 arises from the biexciton comprising approximately two excitons. The information carried by $S(\omega)$ will also be further used to quantify the environmental decoherence. First, we use $S(\omega)$ to demonstrate the presence of parametric multi-phonon processes during state preparation. In \subfigref{fig:evolution}{d}, we show $S(\omega)$ for the one-phonon case, showing peaks only at zero and the Rabi/acoustic frequency. \asubfigref{fig:evolution}{e} and \subfigref{fig:evolution}{f} reveal successive multiples of $\omega_{\text{ac}}$, corresponding to fractions of $\Omega_{\text{R}}$, with additional beat frequencies.

\begin{figure}[tb!]
    \centering
    \includegraphics[width=1\linewidth]{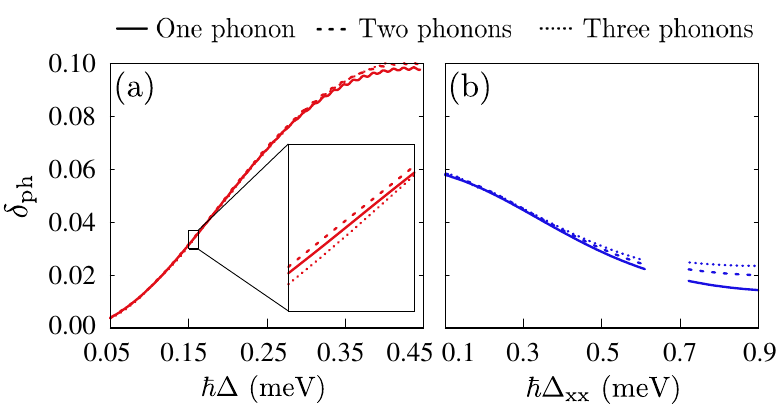}
    \caption{Error of the (a) exciton and (b) biexciton preparation as a function of the corresponding detuning for one- (solid lines), two- (dashed), and three-phonon (dotted) processes for GaAs QDs. The inset in (a) shows minimal differences between curves. In (b), we omit a narrow range of detunings around $0.65$~meV, where we encounter an unwanted resonance leading to exciton occupation.}
    \label{fig:error}
\end{figure}

\subsection{Preparation error}
Besides recombination, subtle dissipative processes arise from interactions with phonons of the host material. Excitons and biexcitons couple to deformations, enabling bath phonons to respond to charge evolution and leading to system-bath entanglement and decoherence of QD states. To address this non-Markovian effect, we include carrier-phonon coupling via the deformation potential and the piezoelectric effect, and employ the second-order Born approximation. We determine infidelity $\delta_{\mathrm{ph}}$ by calculating the final difference in the unperturbed and bath-phonon-impacted density matrices. For this, we follow the approach originally introduced in Refs.~\cite{Alicki2002, Alicki2004} and applied in Refs.~\cite{Roszak2005, Kuniej2025} that allows us to find $\delta_{\mathrm{ph}}$ as an intuitively understandable overlap of $S(\omega)$ and reservoir's spectral density $R(\omega) = |n(\omega) {+} 1|\hbar^{-2}\sum_{\bm{q},\lambda}|g_{\mathrm{x}}^{(\bm{q, \lambda})}|^2\delta(|\omega| {-} \omega_{\bm{q}, \lambda})$, with $n(\omega)$ the Bose-Einstein distribution, that shows frequencies at which the environment can respond (see Supplemental Material for details). The filled curves in \subfigsref{fig:evolution}{d}{f} display $R(\omega)$ calculated for a GaAs QD, demonstrating its overlap with $S(\omega)$. Notably, $R(\omega)$ peaks at $\sim0.8$~meV, beyond the plot range. For large detunings and thus phonon frequencies that exceed this value, almost decoherence-free evolution is possible \cite{Kuniej2025}. Here, we focus on the lowest phonon frequencies to meet the technological limitations in acoustic generation. Thus, we are, in turn, below the $R(\omega)$ maximum, ensuring that the system experiences relatively weak phonon response during evolution.

In \subfigref{fig:error}{a} and \subfigref{fig:error}{b}, we show the exciton and biexciton preparation error at $T=4$~K. The differences between acousto-optical processes of different harmonic orders are minimal, indicating the potential to utilize higher acoustic harmonics. Comparing this result with \subfigref{fig:fractionalMapOccupation}{a}, we predict the potential to prepare an excited state with an error below $2\%$. For the exciton state, the error increases rapidly with detuning until reaching the frequency of maximal phonon response. For the biexciton state, the error decreases with two-photon detuning, as the transition between the exciton and biexciton states is crucial \cite{Kuniej2025}.

\begin{figure}[tb!]
    \centering
    \includegraphics[width=0.94\linewidth]{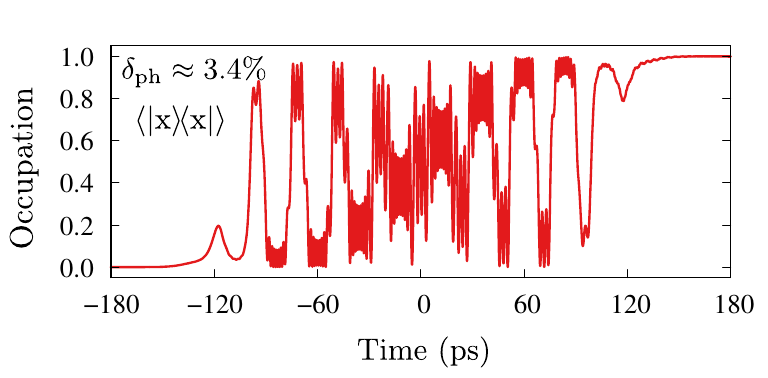}
    \caption{Evolution of exciton occupation in the sub-THz detuning regime. Using 10-phonon processes, QD is excited with an acoustic wave with feasible frequency $\omega_{\mathrm{ac}}/2\pi\approx42$~GHz and $\hbar A_{\mathrm{ac}}\approx2.49$~meV on top of a far-detuned cw laser with $\hbar\Delta = 1$~meV, and $A_{\mathrm{L}} = 1$~meV.}
    \label{fig:subTHzRegime}
\end{figure}

\subsection{Approaching THz dynamics}
Having understood the effect and estimated the achievable fidelity, we now show how parametric multi-phonon dressed-state resonances can bridge over an order of magnitude in frequency, enabling acoustic driving of sub-THz dynamics. As a proof of principle, \figref{fig:subTHzRegime} illustrates that an acoustic pulse at 42~GHz allows the excitation of the exciton via a multiple of $n=10$ for a sub-THz laser detuning of 1~meV and laser amplitude of $1$~meV. These conditions correspond to an optical Rabi frequency of $\Omega_{\mathrm{R}}/2\pi\approx0.341$~THz, surpassing accessible base acoustic frequencies (the ratio of $\Omega_{\mathrm{R}}/\omega_{\mathrm{ac}}$ differs from 10 due to additional splitting of the doubly-dressed levels).
Despite high $n$ and multiple beat frequencies in regions of high phonon spectral density, the preparation error remains low at 3.4\%. This demonstrates that the proposed multi-phonon swing-up scheme may be a practical route toward the THz splitting regime. This example displays how the scheme decouples energy setting from coherent control, with the state splitting addressed by the laser frequency and state manipulation achieved using the lower-frequency acoustic field.

\section{Conclusions and outlook}
We have proposed an extension of the hybrid acousto-optical method for preparing exciton and biexciton states in QDs using higher acoustic frequency multiples.
Even a monochromatic acoustic modulation can effectively generate higher harmonics through multi-phonon-like processes arising from periodic phase modulation of optical transitions, leading to parametric resonances with the Rabi splitting of the laser-dressed exciton.
We have shown deterministic preparation of exciton and biexciton states using multiples of the base acoustic frequency in combination with a detuned laser, achieving low errors that can be further reduced by increasing the detuning. In particular, the method can bridge at least an order of magnitude in frequency, as illustrated by predicting the successful driving of 0.341-THz dynamics with 42-GHz acoustic modulation. This establishes coherent control of optical-QD states while separating optical energy delivery from acoustic control. Beyond driving charge states in QDs, this approach can enable deterministic multi-phonon state generation in acoustic resonators via a detuned laser and facilitate information transfer between solid-state systems using quantized acoustic modes.

\section*{Acknowledgments}
We thank Karolina S{\l}owik and Daniel Groll for helpful discussions. M.~K. acknowledges the support by Grant No. 2023/49/N/ST3/03931 from the National Science Centre (Poland).
M.~G. acknowledges the financing of the MEEDGARD project funded within the QuantERA II Program that has received funding from the European Union's Horizon 2020 research and innovation program under Grant Agreement No. 101017733 and the National Centre for Research and Development, Poland -- project No. QUANTERAII/2/56/MEEDGARD/2024. This work has been supported by a Research Group Linkage Grant of the Alexander von Humboldt-Foundation funded by the German Federal Ministry of Education and Research (BMBF).

\section{Data availability}
The datasets generated during this study and used to produce the figures are publicly available in RepOD \cite{RepOD}.

\bibliography{literature}

\end{document}


\author{Mateusz Kuniej\,\orcidlink{0000-0001-5476-4856}}
\author{Pawe{\l} Machnikowski\,\orcidlink{0000-0003-0349-1725}}
\author{Micha{\l} Gawe{\l}czyk\,\orcidlink{0000-0003-2299-140X}}
\affiliation{Institute of Theoretical Physics, Wroc\l{}aw University of Science and Technology, 50-370 Wroc\l{}aw, Poland}

\title{Supplementary Material:\\Higher-harmonic acoustic control of laser-dressed quantum-dot states at sub-Rabi frequencies}

\begin{abstract}
In this Supplementary Material, we provide details of the fidelity calculations and quantum-dot modeling used in the main paper.
\end{abstract}

\maketitle

\section{Fidelity calculation in general open systems}\label{sec:app-fidelity}
We consider a quantum dot modeled as a three-level system coupled to the acoustic phonon bath. Our goal is to find the fidelity of the prepared state, including the pure-dephasing interaction between carriers and phonons, accounted for beyond the Markov approximation. First, we define the specific interaction Hamiltonian present in our problem. This is followed by the assumption of weak carrier-phonon interaction, which allows us to treat it perturbatively. We find that the final fidelity can be expressed as an overlap between phonon spectral density and the dynamical spectrum of the system.

To assess the fidelity, we follow the approach described in Refs.~\onlinecite{Alicki2002, Alicki2004}, adapted to our problem, as also applied in Refs.~\onlinecite{Roszak2005, Kuniej2025}. The total Hamiltonian of the system is $H_{\mathrm{full}}=H\otimes\mathbb{I}+\mathbb{I}\otimes H_{\mathrm{ph}} + V$. The Hamiltonian of the isolated carrier subsystem is $H = \sum_i E_{i} \ketbra*{E_{i}}{E_{i}}$, $i\in\{\mathrm{g},\mathrm{x},\mathrm{xx}\}$, $H_{\mathrm{ph}}$ is the free phonon Hamiltonian, and $V$ is the coupling.
Without interaction, the evolution of the entire system is
\begin{equation}
    U_0(t) = e^{-iH_{\mathrm{S}}t/\hbar}\otimes e^{-iH_{\mathrm{ph}}t/\hbar},
\end{equation}

Phonons couple diagonally to the exciton and biexciton states \cite{Nazir2016}. We write the coupling in the general form
\begin{equation}
    V = S_{\mathrm{x,x}}\otimes R_{\mathrm{x,x}} + S_{\mathrm{xx,xx}}\otimes R_{\mathrm{xx,xx}}
\end{equation}
where $S_{n,m} = \ketbra{n}{m}$ acts on the carrier subsystem and
\begin{equation}
    R_{n,m} = \sum_{\bm{k}, \lambda}F_{n,m}(\bm{k})\left(b_{\bm{k}, \lambda} + b_{-\bm{k}, \lambda}^{\dagger}\right)
\end{equation}
on the environment with the coupling functions $F_{n,m}(\bm{k}) = F_{m,n}(-\bm{k})$. Here $b_{\bm{k}, \lambda}$ ($b_{\bm{k}, \lambda}^{\dagger}$) is the annihilation (creation) operator for phonons in a mode with wave-vector labeled by $\bm{k}$ and polarization $\lambda$ with three values, LA, TA1, TA2, corresponding to longitudinal and two transverse acoustic branches. Further, we assume $R_{\mathrm{xx,xx}}\approx 2R_{\mathrm{x,x}}$, which allows us to simplify the interaction term as
\begin{equation}
    V = S \otimes R,
\end{equation}
with $S = \ketbra{\mathrm{x}}{\mathrm{x}} + 2\ketbra{\mathrm{xx}}{\mathrm{xx}}$, and $R = R_{\mathrm{x,x}}$.

The evolution of the full density matrix $\varrho(t)$ is described by the Liouville-von Neumann equation
\begin{equation}
    \Dot{\varrho}(t) = -\frac{i}{\hbar}\left[H+ H_{\mathrm{ph}} + V, \varrho(t)\right],
\end{equation}
and we assume a product initial state $\varrho(t_0) = \rho_0 \otimes \rho_{\mathrm{R}}$, where $\rho_{0} = \ketbra*{\psi_0}{\psi_0}$ is the carrier system's initial density matrix, and $\rho_{\mathrm{R}}$ is the phonon thermal density matrix. Partial trace over the environmental degrees of freedom yields the subsystem's reduced density matrix $\rho(t)=\mathrm{Tr}_{\mathrm{R}}\varrho(t)$. Assuming a perturbative environmental impact, we can write
\begin{equation}
    \rho(t) = \rho^{(0)}(t) + \Delta\rho(t),
\end{equation}
where $\Delta\rho(t)$ is the phonon-induced correction to the unperturbed evolution $\rho^{(0)}(t) = U_0(t)\rho_0U_0^{\dagger}(t)$. We calculate fidelity as the overlap of the final density matrices with and without perturbation
\begin{equation}
    \begin{split}
        F^2 = \Tr\rho^{(0)}(\infty)\rho(\infty) &= \bra{\psi_0}U_0^{\dagger}(\infty)\rho(\infty)U_0(\infty)\ket{\psi_0} \\ &= 1 + \bra{\psi_0}\Delta\Tilde{\rho}(\infty)\ket{\psi_0},
        \label{eq:F2}
    \end{split}
\end{equation}
where tilde denotes the interaction picture, $\tilde{O}(t)=U_0^{\dagger}(t) O U_0(t)$, and $\Delta\Tilde{\rho}(\infty)$ is the perturbation calculated in the second-order Born approximation,
\begin{equation}
   \Delta\Tilde{\rho}(t) = -\frac{1}{\hbar^2}\int_{t_0}^{t}\!\!\mathrm{d}\tau\int_{t_0}^{\tau}\!\!\mathrm{d}\tau'\mathrm{Tr}_{\mathrm{R}}\left[\Tilde{V}(\tau), \left[\Tilde{V}(\tau'), \varrho(t_0)\right]\right].
\end{equation}
After expanding commutators and taking the partial trace, one gets
\begin{align}
     \Delta\Tilde{\rho}(t) =
    -\frac{1}{\hbar^2}\!\int_{t_0}^{t}\!\!\!\mathrm{d}\tau\!\int_{t_0}^{\tau}\!\!\!\!\mathrm{d}\tau'[S(\tau), S(\tau')\rho_0]\langle R(\tau)R(\tau')\rangle + \hc
    \label{eq:deltaRho}
\end{align}
To proceed, we switch to the spectral domain and define the phonon spectral density
\begin{equation}
    R(\omega) = \frac{1}{2\pi\hbar^2}\int_{-\infty}^{\infty}\!\!\mathrm{d}t\left\langle R(t)R\right\rangle e^{i\omega t},
\end{equation}
and the nonlinear dynamical spectrum
\begin{equation}
    S(\omega) = \sum_{n}\left\lvert\bra{\psi_0}Y^{\dagger}(\omega)\ket{\psi_n}\right\rvert^2,
\end{equation}
where states $\{\ket*{\psi_n}\}$ span the space orthogonal to the initial state, and $Y(\omega) = \int_{t_0}^{t}\mathrm{d}\tau \tilde{S}(\tau)e^{-i\omega \tau}$. By using those in Eq.~\eqref{eq:deltaRho} and plugging this back into Eq.~\eqref{eq:F2}, we arrive at the fidelity loss expressed as the overlap of the spectral functions
\begin{equation}
    F^2 = 1-\int_{-\infty}^{+\infty}\!\!\mathrm{d}\omega S(\omega)R(\omega),
    \label{eq:fidelity}
\end{equation}
which we evaluate numerically.

\section{Application to semiconductor QD states}
\begin{table}[tbp!]
\begin{ruledtabular}
\begin{tabular}{ccccccccc}
$a_{\mathrm{c}}$ & $a_{\mathrm{v}}$ & $d$ & $c_{\mathrm{LA}}$ & $e_{14}$ & $\varepsilon_{\mathrm{r}}$ & $l_\rho$  & $l_z$ \\
 (eV) & (eV) & (kg/$\mathrm{m}^3$) & (m/s) & (C/$\mathrm{m}^2$) & &  (nm) & (nm) \\ \hline \\[-8pt]
$-6.56$ & $-1.68$ & 4696 & 5590 & $-0.145$ & 11.8 & 9 & 3  \\
\end{tabular}
\end{ruledtabular}
    \caption{Material parameters from Refs.~\onlinecite{Caro2015, Levinshtein1999, Vurgaftman2001} and structural parameters used to calculate phonon spectral densities for the AlGaAs barrier based on linear interpolation between AlAs and GaAs. For sound velocity, the quadratic interpolation formula from Ref.~\onlinecite{Levinshtein1999} is used.}
    \label{tab:material}
\end{table}

For carriers confined in a quantum dot, the phonon spectral density is
\begin{equation}
    R(\omega) = |n(\omega) + 1| \, \frac{1}{\hbar^{2}}\sum_{\bm{q},\lambda}\left\lvert g_{\bm{k, \lambda}}\right\rvert^2\delta\left(|\omega| - \omega_{\bm{k}, \lambda}\right),
\end{equation}
where $n(\omega)$ is the Bose-Einstein distribution, and the exciton-phonon coupling,
\begin{equation}
    g_{\bm{k, \lambda}} = g_{\bm{k, \lambda}}^{(\mathrm{e, DP})} + g_{\bm{k, \lambda}}^{(\mathrm{e, PE})} + g_{\bm{k, \lambda}}^{(\mathrm{h, DP})} + g_{\bm{k, \lambda}}^{(\mathrm{h, PE})},
\end{equation}
includes couplings via the piezoelectric effect (PE) and deformation potential (DP). Assuming linear phonon dispersion $\omega_{\bm{k}, \lambda} = c_{\lambda}k$, where $c_{\lambda}$ is the sound velocity, the couplings are~\cite{Mahan2000}
\begin{subequations}
    \begin{equation}
        g_{\bm{k, \lambda}}^{(\mathrm{e/h, DP})} = \pm\delta_{\lambda, \mathrm{LA}} \, a_{\mathrm{c/v}}\sqrt{\frac{\hbar k}{2d\mathcal{V}c_{\mathrm{\lambda}}}}\mathcal{F}_{\mathrm{e/h}}(\bm{k}),
    \end{equation}
    \begin{equation}
        g_{\bm{k, \lambda}}^{(\mathrm{e/h, PE})} = \pm i\frac{ee_{14}}{\varepsilon_{0}\varepsilon_{r}}\sqrt{\frac{\hbar}{2d\mathcal{V}\omega_{\bm{k},\lambda}}}M_{\lambda}(\hat{\bm{k}}) \mathcal{F}_{\mathrm{e/h}}(\bm{k}),
    \end{equation}
\end{subequations}
where
\begin{equation}
    \mathcal{F}_{\mathrm{e/h}}(\bm{k}) = \int_{\mathbb{R}^3}\mathrm{d}^3\bm{r}\,\varPsi^{*}_{\mathrm{e/h}}(\bm{r})\varPsi_{\mathrm{e/h}}(\bm{r})e^{i\bm{k}\cdot\bm{r}}
\end{equation}
is the form factor with electron/hole ground-state wave functions $\varPsi_{\mathrm{e/h}}(\bm{r})$, and $M_{\lambda}(\hat{\bm{k}}) = 2\hat{k}_x\hat{k}_y(\hat{\bm{e}}_{\bm{k}, \lambda})_{z} + \mathrm{c.p.}$, where $\hat{\bm{e}}_{\bm{k}, \lambda}$ is the phonon unit polarization vector, $\hat{\bm{k}} = \bm{k}/k$ and c.p. stands for a cyclic permutation of indices. Here, $a_{\mathrm{c}}$ ($a_{\mathrm{v}}$) denotes the electron (hole) deformation potential, $\mathcal{V}$ is the phonon mode normalization volume, $d$ is the crystal density, $e_{14}$ is the piezoelectric constant, and $\varepsilon_{\mathrm{r}}$ is the relative permittivity.

For a QD, we utilize the harmonic confinement approximation with Gaussian wave functions 
\begin{equation}
\varPsi_{\mathrm{e/h}}(\bm{r}) = (2\pi)^{-\frac{3}{2}}\frac{1}{l_z l_\rho^2}\exp\left({-\frac{z^2}{2l_z^2}-\frac{\rho^2}{2l_\rho^2}}\right),
\end{equation}
where $\rho=\sqrt{x^2+y^2}$, and $l_z$ and $l_\rho$ are the out-of-plane and in-plane extensions. We assume stronger hole localization compared to the electron, i.e., $l_i^{(\mathrm{h})} = 0.8 l_i^{(\mathrm{e})}$. The material parameters for calculating the phonon spectral densities are listed in Table~\ref{tab:material}.

\bibliography{literature}